# PROVIDENCE: a ground station for space observation

P.-L. Mayeur[1], J.-F. Sauvage[1 2], T. Fusco[1 2 3], N. Védrenne[1], C. Petit[1], D. Barbon-Dubosc[1], G. Leplat[1], G. Guerrin[1], A. Vincent-Randonnier[1], V. Michau[1],
F. Dolon[3], M. Ferrari[3], F. Huppert[3], P.-E. Blanc[2], S. Basa[2 3], J. Floriot[2], B. Neichel[2],
T. Guerini[4], J. Mingot[4], T. Jolivet[4], R. Buchard[4], N. Robbe[4], J. Dezeuze[4],
S. Idri[5], F. Rosso[5],
C. Boyer[6], S. Thomas[7], S. Egron[8] and all the PROVIDENCE Team

*[1]ONERA – The French Aerospace Lab, Salon-de-Provence, France ;*
*[2]LAM, Laboratoire d'Astrophysique de Marseille, France ;*
*[3]OHP, Observatoire de Haute-Provence, France ;*
*[4]EMBASE, France; [5]BETEM, France;*
*[6]TMT Observatory; [7]Rubin Observatory; [8]Iridescence-group*

## ABSTRACT

PROVIDENCE is an ONERA-led project to install a new-generation optical ground station at the Haute-Provence Observatory (OHP) in South of France. Built around a 2.5-meter aperture, the station addresses five core science cases: space domain awareness, astronomy, laser activities and communications, atmospheric characterization, and instrumental prototyping. The project is organized into three parallel subprojects covering the telescope (Providence T), the building (Providence B), and the instruments (Providence I), the latter including the facility's first-light adaptive optics instrument, INTERSTELLAR. This paper presents the scientific motivation, the project architecture, the telescope technical specifications, the civil-engineering challenges associated with replacing the existing T152 equatorial telescope, and the station layout designed to host multiple co-active instrument teams. The project roadmap targets First Light in 2029, timed to track the close-approach flyby of asteroid Apophis.

## 1. INTRODUCTION

ONERA, the French Aerospace Lab, is leading the development of PROVIDENCE, a new-generation optical ground station dedicated to space observation. The project aims to install a state-of-the-art telescope facility at the Haute-Provence Observatory (OHP), in southern France. The acronym PROVIDENCE stands for Platform for Research in Optics, Vector of Innovation for Defense, Environmental understanding and Characterization of Objects in Space. The station is conceived as an optical research platform optimized for high angular resolution, intended to serve both the defense and civilian astronomical communities. This paper presents the scientific motivation behind the project, its architecture, its technical specifications, and its current development status.

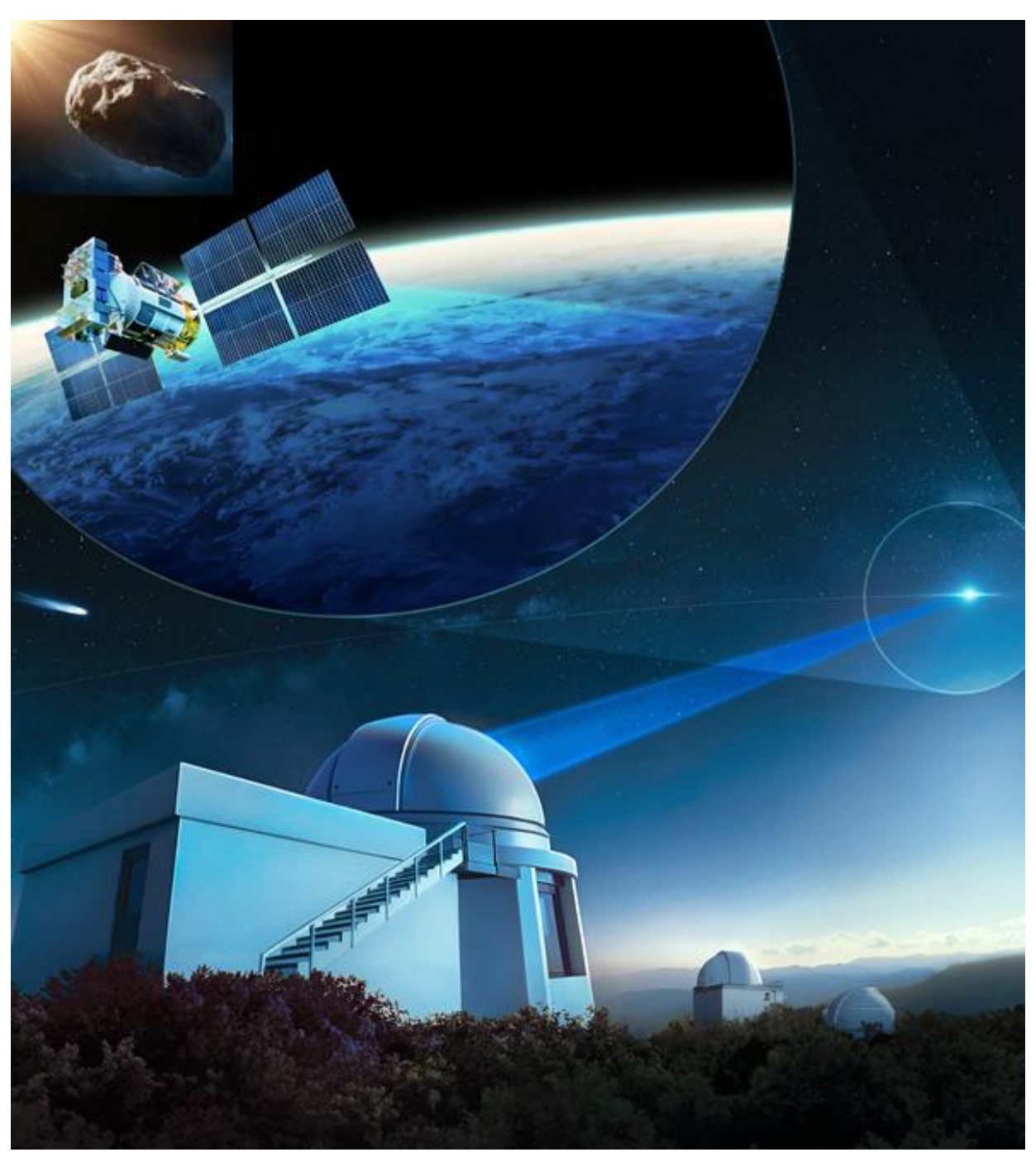

**Figure 1.** Artist's view of the PROVIDENCE optical ground station, its 2.5-meter telescope, and its space observation missions, from satellite tracking in Earth orbit to asteroid monitoring.

## 2. CORE SCIENTIFIC MISSION CASES

ONERA has defined five main science cases for the PROVIDENCE optical ground station, described below. [1, 2]

### 2.1 Space Domain Awareness

This case focuses on high-resolution imaging of objects in Low Earth Orbit (LEO) and Very Low Earth Orbit (VLEO). It also covers deep surveying of the close environment of the Geostationary Earth Orbit (GEO), high-precision orbitography, and space debris monitoring. [3, 4, 5, 6, 7, 8, 9, 10]

### 2.2 Astronomy

This case includes the observation of transient astronomical phenomena, as well as the tracking and monitoring of Solar System bodies such as asteroids, Near-Earth Objects (NEOs), and planets. [

### 2.3 Laser activities and communications

PROVIDENCE will support satellite laser ranging, active object illumination, deep-space laser communications, and Quantum Key Distribution (QKD) protocols.

### 2.4 Atmospheric characterization

This case involves real-time monitoring of atmospheric turbulence optical parameters and transmission properties, as well as structural profiling, such as contrail behavior analysis.

### 2.5 Instrumental prototyping

For all the science cases above, PROVIDENCE will act as a polyvalent station, open to the community for instrumental development, testing, and validation of advanced concepts, in particular for high spatial and spectral resolution imaging.

## 3. PROJECT ARCHITECTURE

The project is structurally divided into three parallel subprojects:

- **Providence T (Telescope):** Covers the telescope system, the dome system, and the control system.
- **Providence B (Building):** Encompasses the civil engineering, building envelope modifications, and pier structural upgrades at OHP.
- **Providence I (Instruments):** Focuses on backend instrumental design, notably the integration of the facility's first-light adaptive optics instrument: INTERSTELLAR.

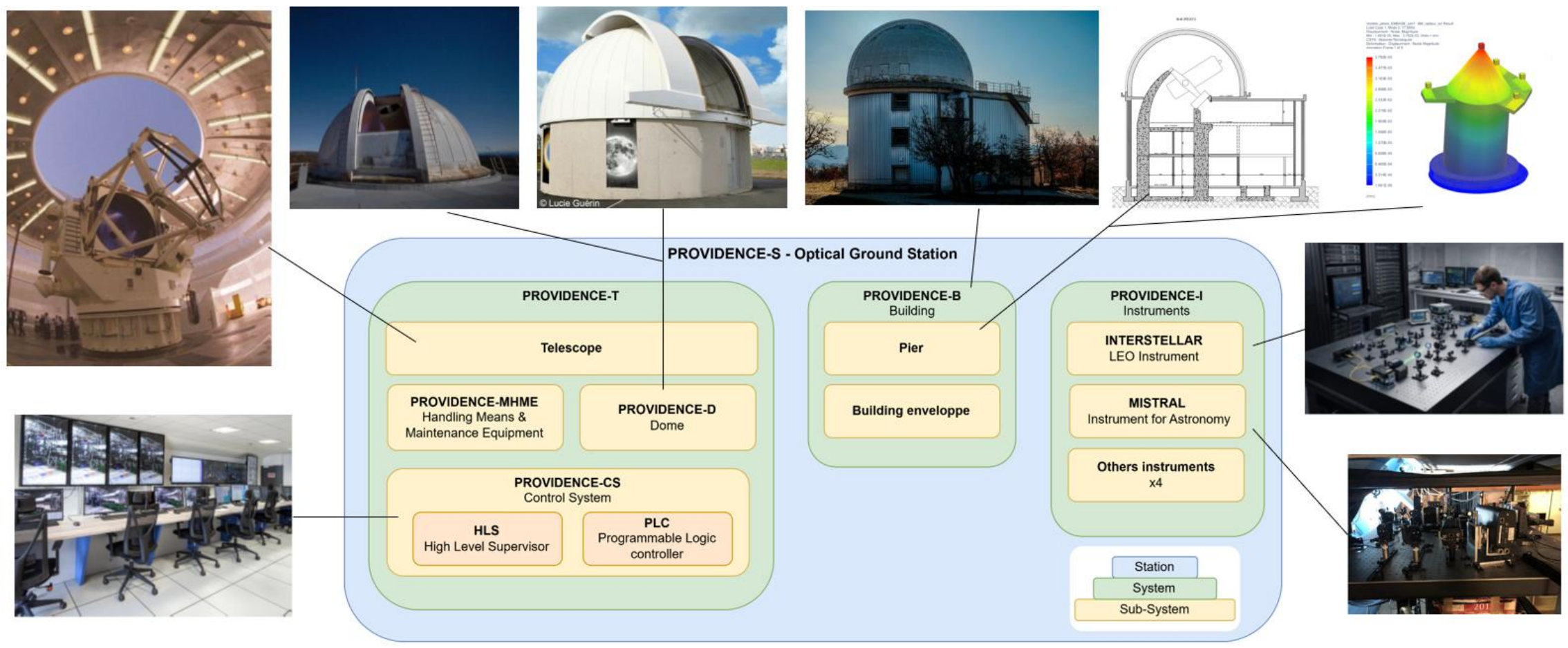

**Figure 2.** PROVIDENCE project architecture, showing the three parallel subprojects Providence T (Telescope), Providence B (Building), and Providence I (Instruments)

## 4. PROJECT ROADMAP AND TIMELINE

The project is fully on track. Table 1 summarizes past milestones and upcoming steps.

| Year | Milestone |
|---|---|
| 2023 | Preliminary studies launched by ONERA to define user needs for the station. |
| 2024 | Official project kick-off; high-level mission specifications. |
| 2025 | Top-level requirements translated into functional and technical specifications for the telescope;<br>Call for Tender for the telescope launched. |
| 2026 | Technical and functional specifications for the building redaction;<br>Call for Tender for facility modernization ongoing.<br>**Project fully funded:** PROVIDENCE is supported by the European Regional Development Fund (ERDF), under the Provence-Alpes-Côte d'Azur Region and Alps Massif / ERDF-ESF+JTF 2021-2027 program, to support projects contributing to the economic, scientific, and technological development of the territory |
| 2027 | On-site construction and building modification work start at OHP. |
| 2028 | Installation and alignment of the telescope mount, optics, and dome assembly. |
| 2029 | First Light, with the primary operational target of tracking the close-approach flyby of asteroid Apophis (April 2029). |

Table 1. PROVIDENCE project roadmap, from the preliminary studies (2023) to First Light (2029).

## 5. TELESCOPE TECHNICAL SPECIFICATIONS (PROVIDENCE T)

The PROVIDENCE telescope represents a major technological leap forward, characterized by the following parameters:

- **Aperture and resolution:** A 2.5-meter primary aperture designed to achieve a spatial resolution of **41mas at 0.5µm (**i.e. 10cm at 500km)
- **Spectral coverage:** a broad operational bandwidth spanning from 380nm (UV) to 2300nm (IR).
- **Field of view:** 10 to 30 arcminutes at the Nasmyth focus and 2arcmin FOV at Coudé focus.
- **Tracking capability:** a high-speed alt-azimuth mount optimized for rapid satellite angular tracking in LEO (up to 5°/s), equipped with a dual-stage tracking loop consisting of a Large Field Seeker (LFS) and a Medium Field Seeker (MFS) to guarantee robust closed-loop tracking.
- **Focal stations:** multi-focal versatility offering two Nasmyth platforms, four Coudé focus stations, and one Cassegrain focus.
- **Control architecture:** fully remote-controllable, designed to evolve into a fully autonomous, robotic system.
- **Laser optimization:** full compatibility with Laser Guide Star (LGS) operations and monostatic high-power laser paths, with the capacity to host up to six auxiliary telescopes on the main mount.

Following a rigorous procurement process, AMOS (Advanced Mechanical and Optical Systems, Belgium) has been selected as the prime contractor for the telescope design and fabrication. [11]

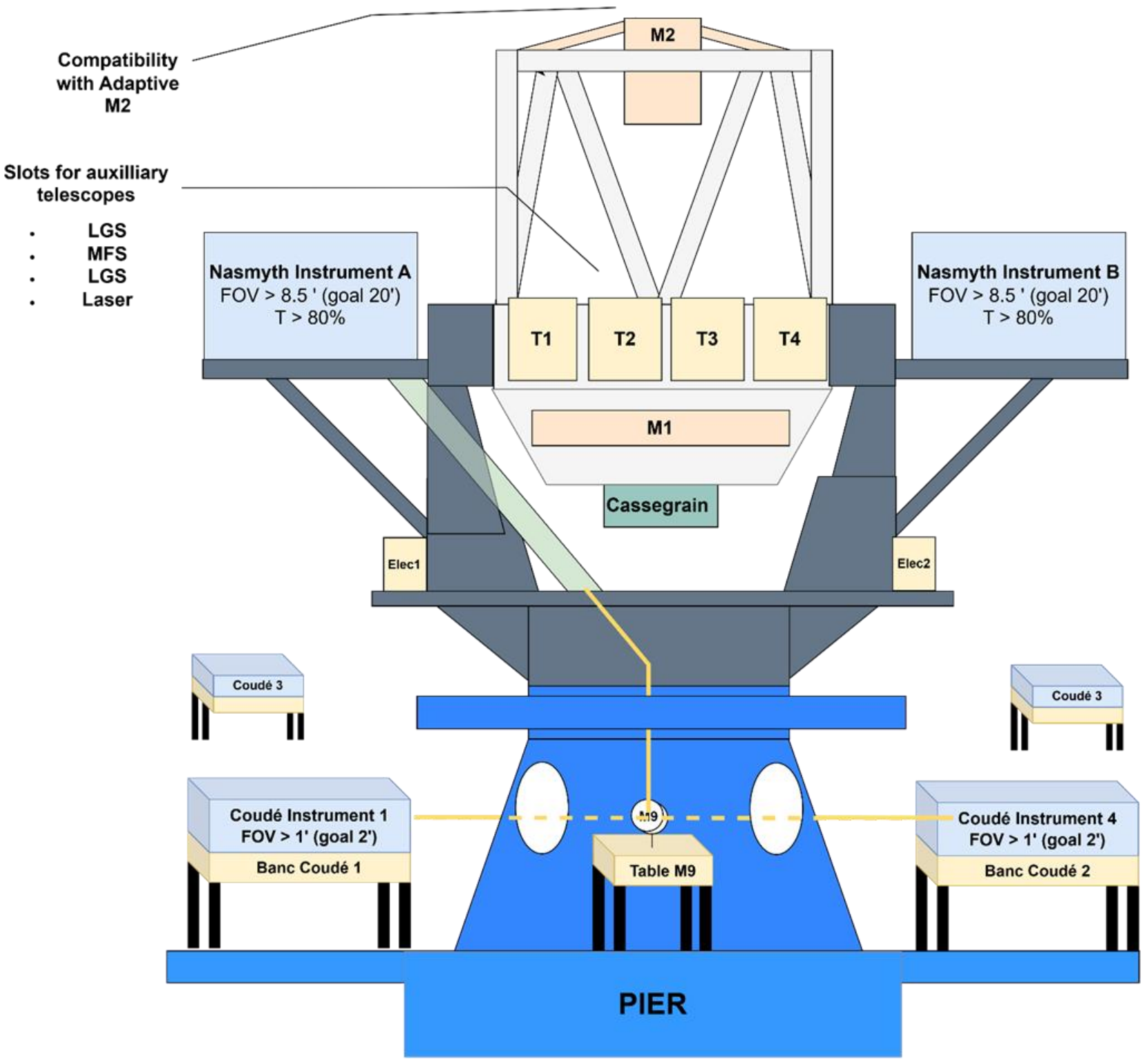

**Figure 3.** PROVIDENCE telescope optical and mechanical architecture

## 6. INFRASTRUCTURE AND BUILDING MODERNIZATION (PROVIDENCE B)

The telescope will be installed in an existing historical building at OHP, which currently houses the T152 equatorial telescope. Replacing this legacy instrument introduces two major civil-engineering challenges.

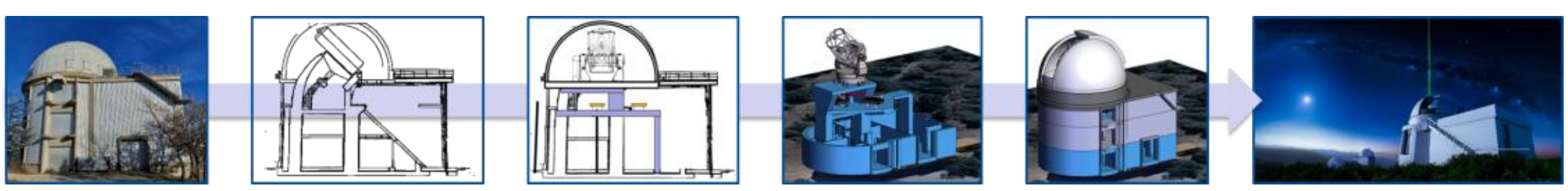

**Figure 4.** Modernization sequence of the historical OHP building, from the current T152 facility to the future PROVIDENCE station

### 6.1 Static load constraints

The current T152 telescope assembly weighs approximately 9 tons. It will be replaced by the PROVIDENCE telescope, which weighs approximately 40 tons. The existing concrete structure and pier must be heavily reinforced or fully replaced to support this substantial increase in static load.

### 6.2 Dynamic load constraints

Tracking fast-moving LEO targets induces severe dynamic torque that must be fully absorbed by the facility's 8-meter-high concrete pier. Strict rigidity and stiffness constraints are mandatory to maintain pointing-model accuracy and tracking performance, together with strict vibration-isolation criteria to preserve the high angular resolution. Preliminary structural and soil-mechanics studies are currently underway to validate the feasibility of the pier modernization.

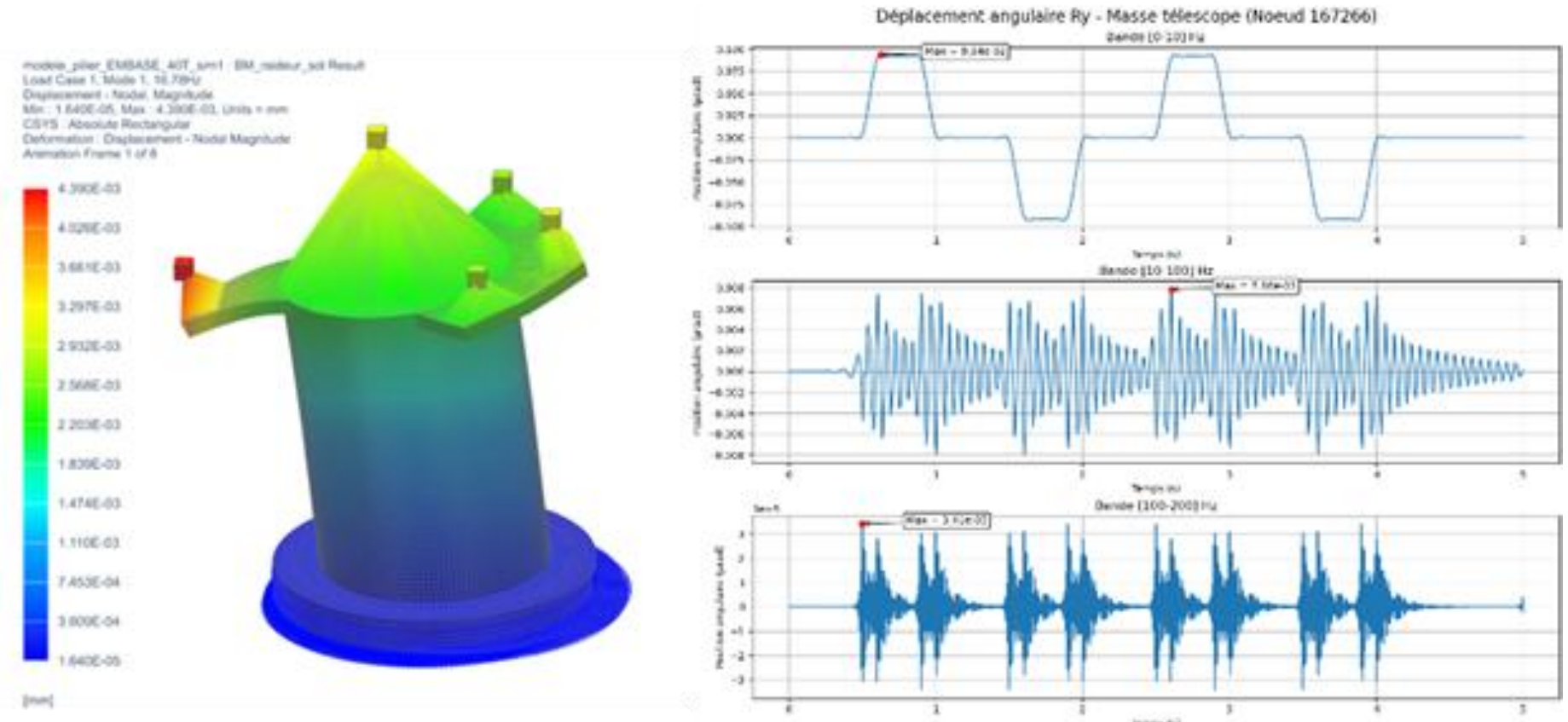

**Figure 5.** Preliminary finite-element study of the PROVIDENCE pier: angular displacement mode shape of the telescope mass (left) and simulated angular response in three frequency bands (0-10 Hz, 10-100 Hz, 100-200 Hz), used to validate the pier's rigidity and vibration-isolation performance.

## 7. STATION LAYOUT AND INSTRUMENTAL ACCOMMODATION

PROVIDENCE is designed to maximize operational efficiency and permit smooth co-activity between separate research teams, through a highly versatile focal-station layout.

- **Nasmyth platforms (×2):** configured to accommodate large-scale instruments weighing up to 500 kg within a 1.5 $m^3$ volume envelope, featuring dedicated water-cooling loops and optimized for automated data-production pipelines.
- **Cassegrain focus (×1):** optimized for compact auxiliary instruments up to 100 kg.
- **Coudé rooms (×4):** the four Coudé foci feed independent laboratories tailored to specific operational requirements: two conventional Coudé labs dedicated to standard optical and astronomical testing; one Defense and High-Power Laser Lab, a highly secure environment with restricted physical access and air-gapped networking designed for high-energy laser transmission and sensitive defense applications;

and one Prototyping and Teaching Lab, featuring an extended floor surface area to accommodate larger groups, students, and temporary experimental setups for rapid instrument prototyping.

To facilitate on-site Assembly, Integration, and Testing (AIT), the station includes dedicated optical, electronic, and mechanical workshops. The control-room architecture is split into one telescope control room and two independent instrument control rooms, allowing multiple teams to operate separate instruments simultaneously during the same observation night without mutual interference.

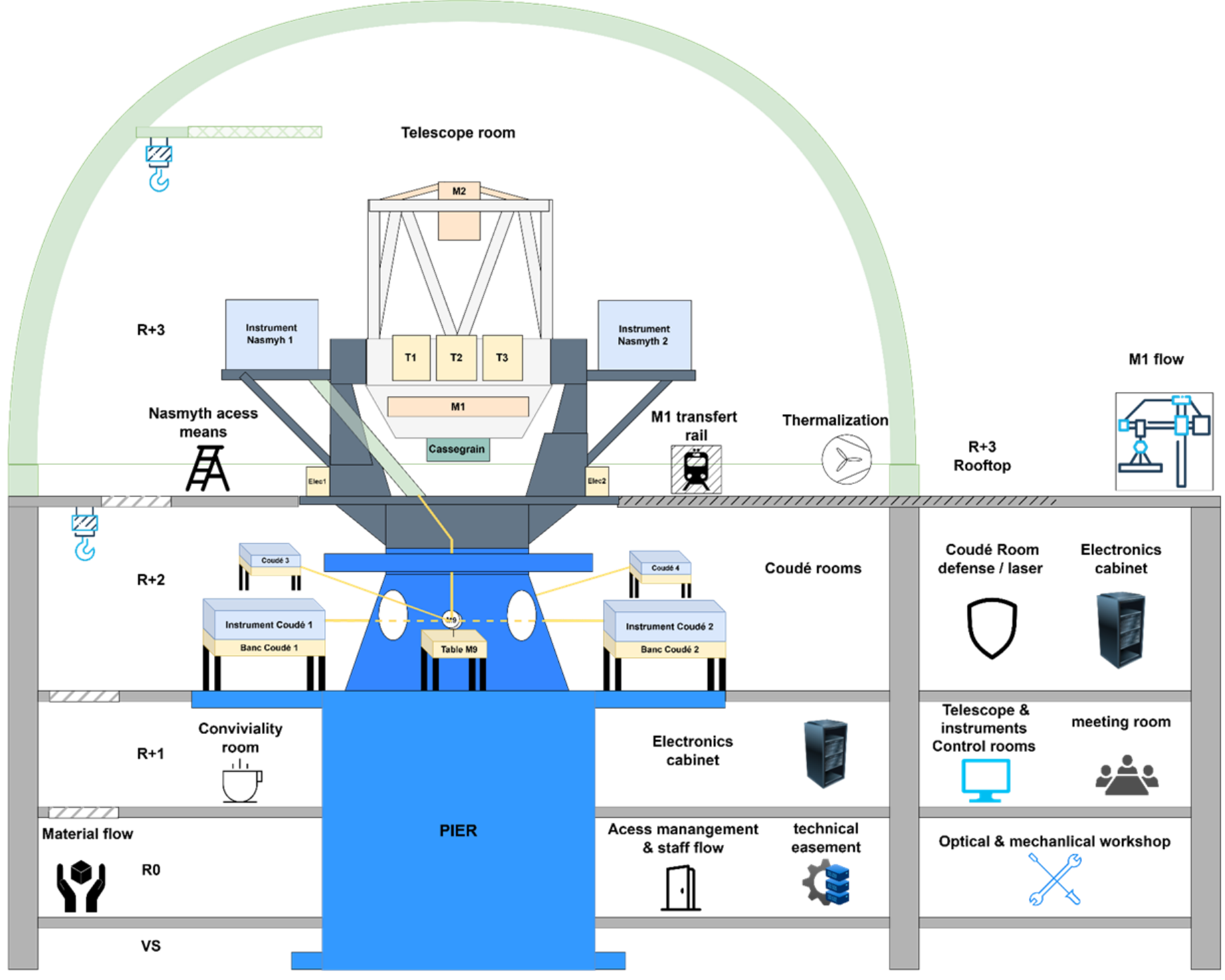

**Figure 6.** Functional layout of the PROVIDENCE station across its vertical levels, from the telescope room and Coudé rooms to the control rooms, workshops, and technical facilities.

## 8. CONCLUSION

PROVIDENCE will provide the European astronomical and defense communities with an unmatched asset for high-angular-resolution space observations. By blending over three decades of ONERA expertise in adaptive optics with an open-access platform philosophy, the station will deliver high-duty-cycle, on-demand observation capabilities. The project is ready for First Light in 2029, with a 2.5-meter telescope and its AO-assisted instrumentation. ONERA actively welcomes international partnerships and collaborative frameworks around this facility.

## ACKNOWLEDGMENTS

The author thanks the ONERA PROVIDENCE project team, the Haute-Provence Observatory, EMBASE and BETEM for their ongoing contribution to the project. The PROVIDENCE infrastructure is co-funded by the European Regional Development Fund (ERDF) under the operational program Provence-Alpes-Côte d'Azur Region and Alpine Massif / ERDF-ESF+-JTF 2021-2027.